\documentclass[useAMS,usenatbib]{mn2e}

\usepackage{times}
\usepackage{graphicx}
\usepackage{txfonts}
\newcommand{\age}{$t$}
\newcommand{\metal}{[$Z$/H]}
\newcommand{\abund}{[E/Fe]}
\newcommand{\agessp}{$t_{\rm SSP}$}
\newcommand{\metalssp}{[$Z$/H]$_{\rm SSP}$}
\newcommand{\abundssp}{[E/Fe]$_{\rm SSP}$}
\newcommand{\agel}{$t_{\rm LW}$}
\newcommand{\metall}{[$Z$/H]$_{\rm LW}$}
\newcommand{\abundl}{[E/Fe]$_{\rm LW}$}

\title[Interpretation of SSP-equivalent parameters]{On the interpretation of age and chemical composition of composite stellar populations determined with line-strength indices.}
\author[P. Serra and S.C. Trager]{P. Serra\thanks{E-mail: pserra@astro.rug.nl} and S.C. Trager\\
Kapteyn Astronomical Institute, RuG, Landleven 12, 9747 AD, Groningen, NL}
\begin{document}

\date{6 September 2006}

\pagerange{\pageref{firstpage}--\pageref{lastpage}} \pubyear{2002}

\maketitle

\label{firstpage}

\begin{abstract}

We study the simple-stellar-population-equivalent (SSP-equivalent) age and chemical composition measured from the Lick/IDS line-strength indices of composite stellar populations (CSP). We build two sets of $\sim$30000 CSP models using stellar populations synthesis models, combining an old population and a young population with a range of ages and chemical compositions representative of early-type galaxies. We investigate how the SSP-equivalent stellar parameters of the CSP's depend on the stellar parameters of the two input populations; how they depend on $V$-band luminosity-weighted stellar parameters; and how SSP-equivalent parameters derived from different Balmer-line indices can be used to reveal the presence of a young population on top of an old one. We find that the SSP-equivalent age depends primarily on the age of the young population and on the mass fraction of the two populations, and that the SSP-equivalent chemical composition depends mainly on the chemical composition of the old population. Furthermore, while the SSP-equivalent chemical composition tracks quite closely the $V$-band luminosity weighted one, the SSP-equivalent age does not and is strongly biased towards the age of the young population. In this bias the age of the young population and the mass fraction between old and young population are degenerate. Finally, assuming typical error bars, we find that a discrepancy between the SSP-equivalent parameters determined with different Balmer-line indices can reveal the presence of a young stellar population on top of an old one as long as the age of the young population is less than $\sim$2.5 Gyr and the mass fraction of young to old population is between 1\% and 10\%. Such disrepancy is larger at supersolar metallicities.
\end{abstract}


\section{Introduction} 
\label{intro}

The knowledge of age and chemical composition of stars in early-type galaxies is a fundamental piece in the puzzle of galaxy formation and evolution. For a long time, optical-wavelength studies in this direction have been hampered by the age-metallicity degeneracy: an age variation of a factor of $\sim$3 mimics a metallicity variation of a factor of $\sim$2 in the spectra of old stellar populations (e.g., Faber 1973; O'Connell 1986; Worthey 1994). The effort of various authors during the past two decades culminated in the work of Worthey (1994), who showed that age and metallicity can be disentangled by the joint use of pairs of line-strength indices, one metal-line and one Balmer-line index, measured from the optical spectra of galaxies. A system of line-strength indices was defined (the Lick/IDS system, see Burstein et al. 1984; Worthey et al. 1994 and references therein; Worthey \& Ottaviani 1997) and is now widely used in order to determine the age \age,  metallicity \metal\ and abundance ratio \abund\ of stars in galaxies (\abund\ is defined in Trager et al.\ 2000a as a way of parameterising deviations from the solar abundance pattern).

In practise, one compares the indices measured from the optical spectrum of a galaxy to their values predicted by stellar populations models (provided for example by Worthey 1994; Vazdekis 1999; Bruzual \& Charlot 2003; Thomas, Maraston \& Bender 2003). The stellar \age, \metal\ and \abund\ of the galaxy are the ones of the model whose indices best agree with the measured ones. Because each model consists of a single-burst stellar population (SSP) whose stars, unlike in real galaxies, all have the same \age, \metal\ and \abund, the derived stellar parameters are labelled as \emph{SSP-equivalent}. We will refer to them as \agessp, \metalssp\ and \abundssp.

Early-type galaxies are the most massive stellar systems for which the SSP approximation seems to hold. Therefore, many authors have measured their line-strength indices in order to determine their stellar content (see for example Trager et al.\ 2000b; Caldwell, Rose \& Concannon 2003; Denicol\'{o} et al.\ 2005; Thomas et al.\ 2005; Clemens et al.\ 2006). However, many results suggest that recent star formation occurred in these galaxies (Trager et al.\ 2000b; Yi et al.\ 2005), so that a small fraction of their current stellar mass formed few Gyr ago. It is natural to wonder about the meaning of the line-strength indices analysis under such circumstances (i.e., in the presence of more than one SSP in the same galaxy). How do the SSP-equivalent parameters relate to the average properties of a galaxy and to the ones of the many SSP's that it hosts? And what do we actually learn from SSP-equivalent parameters?

Previous authors have already looked into this problem. Using a limited number of models of composite stellar populations (CSP), Trager et al.\ (2000b) found that the SSP-equivalent age is heavily biased towards the age of the young stars present in a galaxy. In this paper we address the same questions in a more systematic and extensive way from the point of view of the stellar population models.

We build two sets of CSP's by using the models of Bruzual \& Charlot (2003, hereafter BC03) and Worthey (1994, hereafter W94). Each dataset contains $\sim$3$\times10^4$ CSP models composed of one old and one young SSP. Different CSP's correspond to different stellar parameters of the parent SSP's. The old SSP (SSP$_1$) is always chosen to be more massive than the young one (SSP$_2$), as inferred in many early-type galaxies (e.g., Trager et al.\ 2000b; Leonardi \& Worthey 2000; Jeong et al.\ 2006). We analyse the line-strength indices of the CSP's and derive the corresponding \agessp, \metalssp\ and \abundssp\ as is usually done for observed galaxies. We then analyse how the result depends on the input parameters (\age$_1$, \metal$_1$, \abund$_1$, \age$_2$, \metal$_2$,\abund$_2$ and $\mu$=$M_2$/$M_1$ where $M$ is the stellar mass) and on the luminosity-weighted properties of the CSP's. We restrict our study to systems composed of only two SSP's because this case is still relatively easy to treat and might be a reasonable first-order approximation for early-type galaxies. In Sect.\ref{models} we explain in some details the construction of the two datasets, we present the results in Sect.\ref{results}, and finally draw some conclusions.

\section{Models of composite stellar populations}
\label{models}

\begin{table}
\begin{center}
\caption{Input parameters for the composite stellar populations}
\label{input}
\begin{tabular}{l|l}
\hline
\hline
\noalign{\smallskip}
parameter         & values \\
\noalign{\smallskip}
\hline
\noalign{\smallskip}
\age$_1$ (Gyr)    & 10.0,13.0 \\
\metal$_1$        & $-$0.15,0.0,0.15,0.3 \\
\abund$_1$        & 0.0,0.15,0.3,0.45 \\
\age$_2$ (Gyr)    & 1,1.4,1.8,2.5,3.4,4.7 \\
\metal$_2$        & 0.0,0.15,0.3,0.45 \\
\abund$_2$        & $-$0.15,0,0.15,0.3 \\
$\mu$=$M_2$/$M_1$ & 0.0,0.001,0.005,0.01,0.025,0.05,0.1,0.2,0.35,0.5 \\
\noalign{\smallskip}
\hline
\hline
\end{tabular}
\end{center}

Stellar parameters are labelled as ``1'' and ``2'' corresponding to the old population (SSP$_1$) and the young population (SSP$_2$) respectively. The last row lists the values adopted for the mass fraction between the two populations.
\end{table}

We define a set of old and young SSP's in Table \ref{input}. All the possible combinations of one old and one young SSP for each of the values of the mass fraction $\mu$ give the 30720 CSP models that form each of the datasets. A dataset consists of a 7-dimensional grid in the parameters (\age$_1$, \metal$_1$, \abund$_1$, \age$_2$, \metal$_2$,\abund$_2$, $\mu$). At each position in the grid we store a vector that contains the line-strength indices of the two input SSP's and of the resulting CSP, the luminosity-weighted stellar parameters (\agel, \metall, \abundl) and the SSP-equivalent stellar parameters (\agessp, \metalssp,\abundssp).

The luminosity-weighted stellar parameters are defined in terms of the $V$-luminosity of the parent populations at the respective ages (i.e., taking into account the change in mass-to-light ratio as a population ages).

\begin{figure}
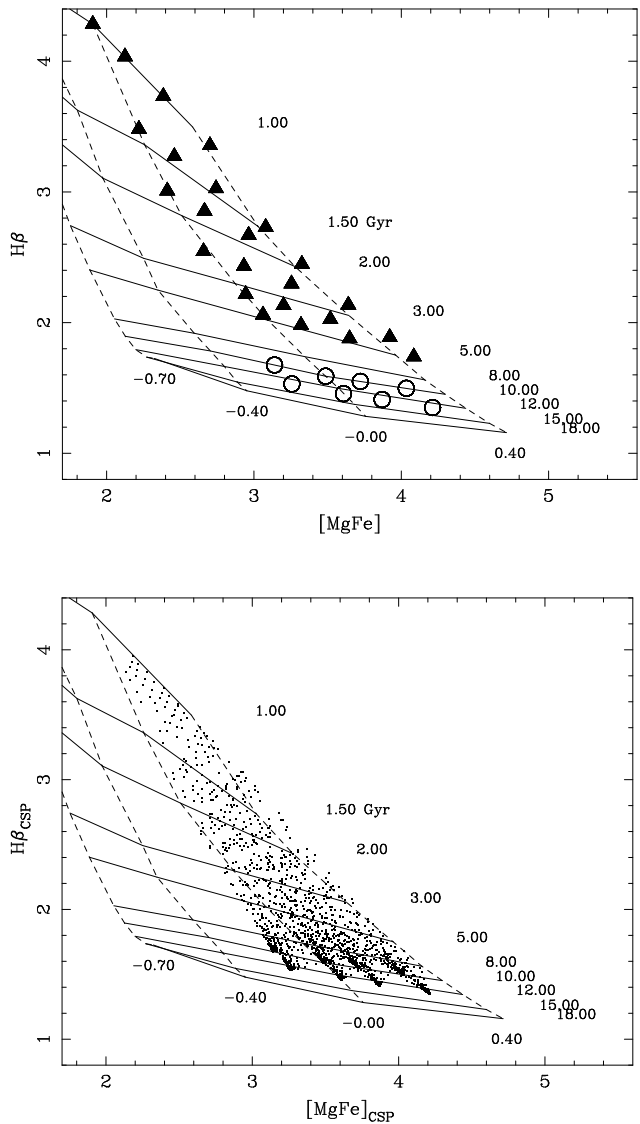

\includegraphics[width=7cm,angle=270]{BC03_solarsample_in.eps}

\vspace{0.8cm}

\includegraphics[width=7cm,angle=270]{BC03_solarsample_out.eps}
\caption{\emph{Top panel}: BC03 parent SSP's of solar \abund\ plotted on top of BC03 model grid; filled triangles represent the young SSP's (SSP$_2$); empty circles represent the old SSP's (SSP$_1$) which dominate the mass of the CSP's (see Table \ref{input}). \emph{Bottom panel}: distribution of the BC03 CSP's built from the solar-\abund\ SSP's.}
\label{solarsample}
\end{figure}

The SSP-equivalent parameters are determined by comparing the Lick/IDS line-strength indices Mg$b$, Fe5270 and Fe5335 and a Balmer-line index of the CSP to their values according to models (BC03 or W94 depending on the dataset). As Balmer-line index we use alternatively H$\beta$, H$\gamma_A$, H$\gamma_F$, H$\delta_A$ or H$\delta_F$, obtaining SSP-equivalent parameters for each of them separately. These will then be labelled according to the Balmer-line index used in deriving them (e.g., \age$_{H\beta}$ is the SSP-equivalent age derived using Mg$b$, Fe5270 and Fe5335 and H$\beta$). The use of different Balmer lines is important because they respond differently to the presence of a young stellar component on top of the old one (Schiavon et al.\ 2004).

\begin{table*}
\begin{center}
\caption{Covariance coefficients between SSP-equivalent parameters and input parameters}
\label{covariance}
\begin{tabular}{cccccccc}
\hline
\hline
\noalign{\smallskip}
\noalign{\smallskip}
          & $\log$ \age$_1$ & $\log$ \age$_2$ & \metal$_1$ & \metal$_2$ & \abund$_1$ & \abund$_2$ & $\mu$ \\
\noalign{\smallskip}
\hline
\noalign{\smallskip}
$\log$ \agessp  & 0.05 , 0.10 &  \emph{0.51 , 0.56} & $-$0.14 , $-$0.07 &  0.09 ,  0.12 &  $-$0.01 ,  0.01 & $-$0.01 ,  0.00 & \emph{$-$0.57 , $-$0.65} \\
     \metalssp  & 0.02 , 0.06 & $-$0.19, $-$0.32 &  \emph{0.78 ,  0.85} &  0.09 ,  0.16 &  $-$0.01 ,  0.01 &  0.00 ,  0.01 &  0.25         \\
     \abundssp  & 0.00 , 0.01 & $-$0.11, $-$0.01 &  0.02 ,  0.07 & $-$0.06 , $-$0.04 &   \emph{0.85 ,  0.87} &  0.23         & $-$0.19 , $-$0.17 \\
\noalign{\smallskip}
\hline
\hline
\end{tabular}
\end{center}
Each entry is the range within which the covariance coefficient varies when changing Balmer-line index (or the value of the coefficient if this does not vary). The largest coefficients for each SSP-equivalent parameter are given in italics.
\end{table*}

\begin{figure}
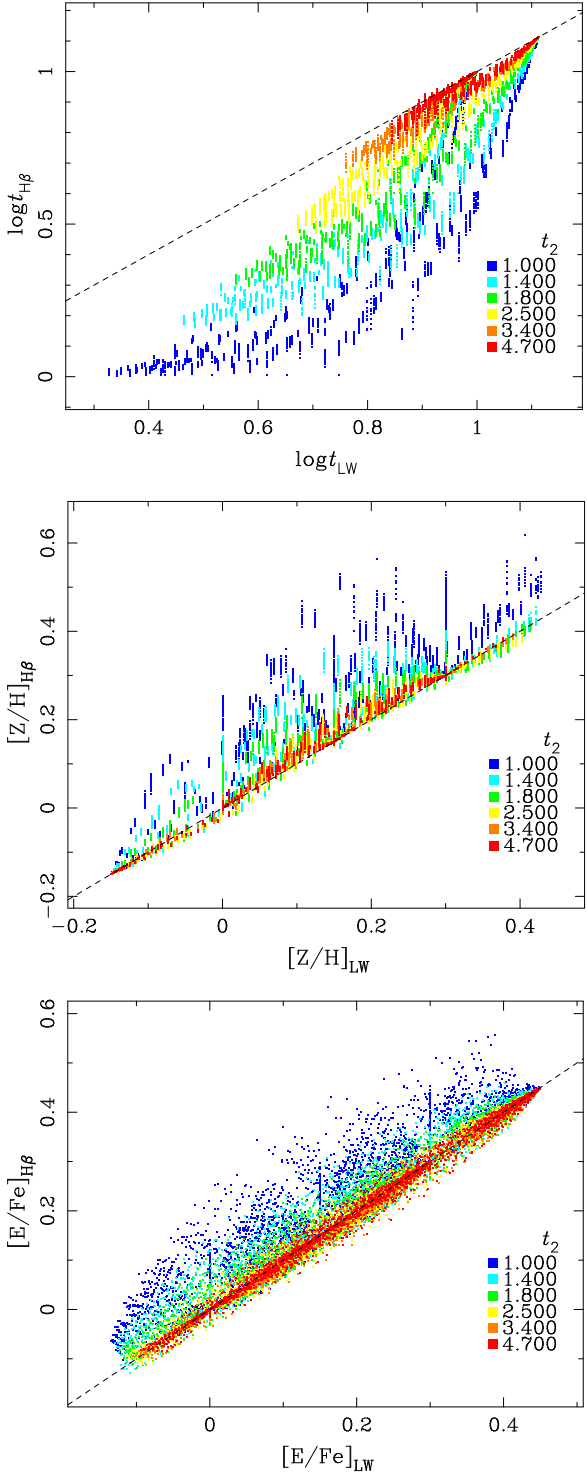


\includegraphics[width=6.3cm,angle=270]{tpaper2.eps}

\vspace{0.3cm}

\includegraphics[width=6.3cm,angle=270]{zpaper2.eps}

\vspace{0.3cm}

\includegraphics[width=6.3cm,angle=270]{epaper2.eps}
\caption{H$\beta$-based \agessp, \metalssp\ and \abundssp\ plotted versus the respective $V$-band luminosity-weighted quantities using BC03 dataset. The dashed line of each plot is the identity line. The colour codes \age$_2$.}
\label{z&e}
\end{figure}

Although W94 and BC03 models are available only with \abund=0, Table \ref{input} contains also parent SSP's of non-solar \abund. In these cases we correct the line-strength indices given by the models according to the \abund\ variations. We then use the corrected values when both building the CSP models and measuring their SSP-equivalent parameters. The correction scheme is the one described in Trager et al.\ (2000a) but improved by the use of new response functions computed and kindly provided by G. Worthey. For details see Trager, Faber \& Dressler (2006).

Fig.\ref{solarsample} shows the distribution of the solar-\abund\ parent SSP's drawn from the BC03 models and of the resulting CSP's on the plane [[MgFe],H$\beta$], where age and metallicity are efficiently decoupled. The points are plotted on top of the BC03 solar-\abund\ model grid. The CSP models cover most of the area where early-type galaxies have been observed to lie (e.g., Trager et al.\ 2000a; Denicol\'{o} et al.\ 2005; Thomas et al.\ 2005).

\section{Results}
\label{results}

We calculate the covariance coefficients between the SSP-equivalent stellar parameters and the input parameters in order to understand which input parameters are mostly driving the variations in \agessp, \metalssp\ and \abundssp. Table \ref{covariance} shows the result of this calculation performed on the BC03 dataset. The covariance coefficients do not change much when changing Balmer-line index. In the table we give the range within which each coefficient varies when changing Balmer-line index. Furthermore, we have verified that the W94 dataset gives the same result. The following comments apply therefore to all Balmer-line indices and to both datasets.

\begin{itemize}

\item The variations in \agessp\ are mostly driven by variations in \age$_2$, the age of the young population, and $\mu$, the mass fraction, while other parameters play a secondary role. The sign and absolute value of these two covariance coefficients clearly show the strong degeneracy between \age$_2$ and $\mu$: the same \agessp\ can result from a small mass of young stars or a sufficiently large mass of older stars.

\item The variations in \metalssp\ are by far dominated by variations in \metal$_1$, the metallicity of the old population. The mass fraction $\mu$ and the age of the young population \age$_2$\ also play a relevant role with the usual degeneracy. The latter correlations must be (at least partially) due to the fact that in the dataset \metal$_2$ is on average larger than \metal$_1$. However, we have verified that the covariance coefficients between \metalssp\ and $\mu$ and between \metalssp\ and \age$_2$ remain significantly larger than zero when considering a subset of models with \metal$_1$ and \metal$_2$ sampled in identical ways (in particular, the covariance coefficients drop by a factor of $\sim$2 and $\sim$1.3 respectively).

\item \abundssp\ seems to be varying mostly because of variations in the abundance ratios of the two parent populations, with the old, massive population being dominant. As for \metalssp, the correlation with $\mu$ and \age$_2$ is only in part due to the different sampling of \abund$_1$ and \abund$_2$. Using a subset of models with identical sampling of \abund$_1$ and \abund$_2$ reduces the covariance coefficient with $\mu$ and increases the one with \age$_2$ by a factor of $\sim$3.

\end{itemize}

\begin{figure}

\includegraphics[width=8cm,angle=0]{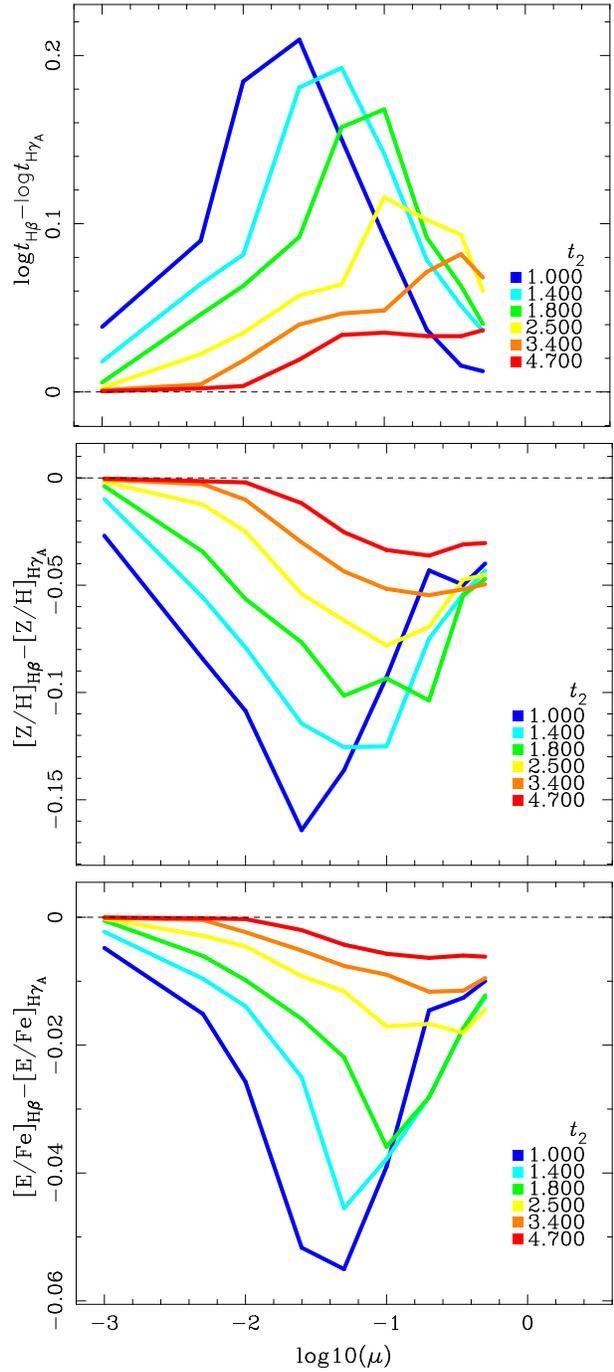}
\caption{Difference between H$\beta$- and H$\gamma_A$-based SSP-equivalent parameters as a function of $\mu$. The plots are obtained using the BC03 dataset and solar \metal\ and \abund\ for both SSP$_1$ and SSP$_2$. The age of SSP$_1$ is fixed at 13 Gyr.}
\label{differences}
\end{figure}

Covariance coefficients highlight which of the input parameters play the dominant role in determining the variation of the SSP-equivalent ones. It is also interesting to see how the latter relate to the average properties of the model CSP's. Fig.\ref{z&e} shows the comparison between the H$\beta$-based SSP-equivalent parameters and the $V$-band luminosity-weighted ones. The behaviour is substantially the same when using different Balmer-line indices.

\metalssp\ and \abundssp\ seem to track quite closely \metall\ and \abundl\ respectively; strong deviations are observed only for the youngest \age$_2$. On the other hand \agessp\ is always much smaller than \agel\ and lies somewhere between the latter and \age$_2$. This effect was already known (Trager et al.\ 2000b). Its explanation is that the determination of \agessp\ relies primarily on Balmer-line indices (see the model grid in Fig.\ref{input}). These are dominated by young stars and therefore \agessp\ is strongly biased towards the age of the young stellar component. As Fig.\ref{z&e} illustrates, the younger SSP$_2$ the stronger this bias is.

Fig.\ref{z&e} demonstrates that it is not correct to use \agessp\ as an estimate of when a galaxy formed its stars (yet, this is often done; see for example Clemens et al.\ 2006). A fair statement would be that \emph{\agessp\ is a Balmer-line-weighted age} and it should always be kept in mind that such age is strongly biased towards the age of young stellar components. Furthermore, as highlighted by the covariance coefficient and confirmed by Fig.\ref{z&e}, the effect of \age$_2$ and $\mu$ on \agessp\ is degenerate. An increasingly older SSP$_2$ can produce the same \agessp\ as long as $\mu$ is properly increased (in Fig.\ref{z&e} $\mu$ increases towards decreasing \agessp\ and \agel).

As mentioned, SSP-equivalent parameters derived from different Balmer-line indices behave substantially in the same way (i.e., Fig.\ref{z&e} looks roughly the same for all of them). However, different Balmer-line indices are sensitive to the presence of young stars at different levels (Schiavon et al.\ 2004). Because of this \agessp, \metalssp\ and \abundssp\ (of a CSP) computed with different Balmer-line indices will not be in agreement. Fig.\ref{differences} illustrates this concept for a subset of the CSP models chosen to have solar chemical composition and \age$_1$=13 Gyr. It can also be seen that the difference between H$\beta$- and H$\gamma_A$-based SSP-equivalent parameters goes to zero for $\mu$ approaching 0 and 1 and peaks between these two extremes at a position dependent on \age$_2$. Furthermore, Fig.\ref{differences} shows once more the degeneracy between \age$_2$ and $\mu$. The same difference between, for example, \age$_{H\beta}$ and \age$_{H\gamma_A}$ can be caused by increasingly older SSP$_2$'s as long as the mass fraction $\mu$ is sufficiently increased.

\begin{figure}

\includegraphics[width=8cm,angle=0]{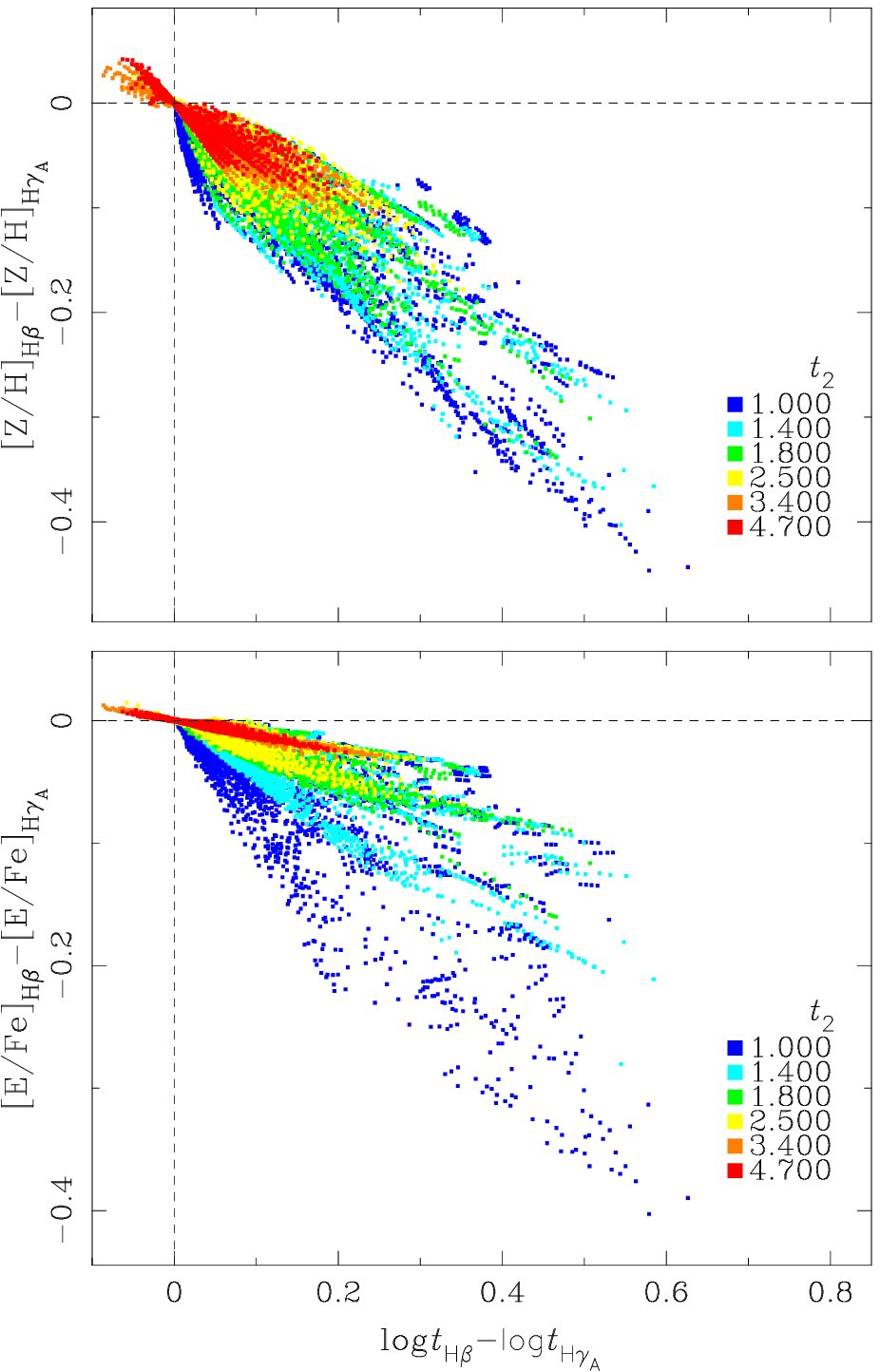}

\caption{Difference between H$\beta$- and H$\gamma_A$-based \metalssp\ (top) and \abundssp\ (bottom) plotted versus the difference in \agessp. Each point corresponds to a BC03 CSP model. The presence of a young stellar population on top of an old one causes SSP-equivalent parameters based on different Balmer-line indices to disagree. This effect must be and indeed is observable simultaneously in \agessp, \metalssp\ and \abundssp. In particular, points seem to be distributed along a very tight relation in the age-metallicity plane.}
\label{totaldeltas}
\end{figure}

For clarity Fig.\ref{differences} shows only a subset of the CSP models. A similar trend is anyway observed in the whole sample (and in W94 dataset), showing that it is possible to detect the presence of a young stellar component on the basis of the disagreement between SSP-equivalent parameters obtained with different Balmer-line indices. We actually expect that more dramatic disagreements in, for example, \agessp\ are accompanied by larger differences in \metalssp\ and \abundssp. This is indeed observed and showed in Fig.\ref{totaldeltas}, where the difference between H$\beta$- and H$\gamma_A$-based \metalssp\ and \abundssp\ is plotted versus the difference in \agessp. In particular the age-metallicity plot shows a very tight relation. This could be used in order to test the correctness of one's results.

Different Balmer lines can be used as a diagnostic for the presence of a young stellar component only as long as the differences in the SSP-equivalent parameters are larger than the observational errors. These are typically of 0.1 on the logarithm of \agessp\ and on \metalssp\ and of 0.05 on \abundssp\ (see for example Trager et al.\ 2000a; Thomas et al.\ 2005). Fig.\ref{totaldeltas} shows that with these errors \agessp\ measurements are the most efficient in revealing a young component, allowing the detection of SSP$_2$'s younger than $\sim$2.5 Gyr. As suggested by Fig.\ref{differences}, this is however possible only within a certain range of $\mu$. The actual range depends on \age$_2$ but we find it to be roughly between 1\% and 10\%. \metalssp\ and \abundssp\ are only sensitive to SSP$_2$'s younger than $\sim$1.5 Gyr with $\mu$ between 2\% and 10\%. It is important to stress that for $\mu\geq$10\% there is no detectable difference between the SSP-equivalent parameters derived from different Balmer-line indices. At these values of $\mu$ the Balmer-line indices are so heavily dominated by the younger populations that they all ``see'' the same age, which is very close to the age of the young population.

\begin{figure}

\includegraphics[width=6.5cm,angle=270]{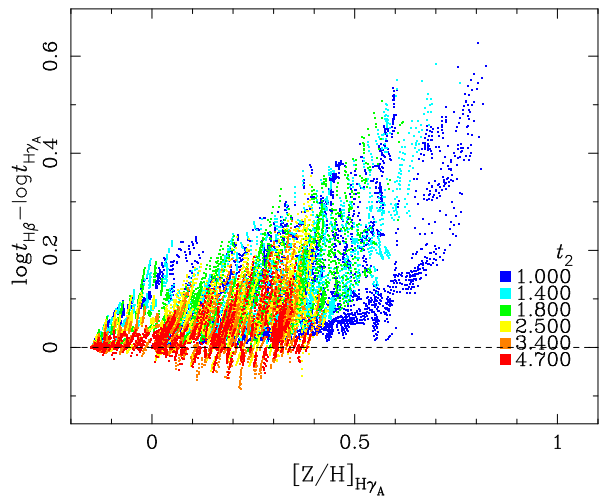}

\caption{Difference between H$\beta$- and H$\gamma_A$-based \agessp\ plotted versus the H$\gamma_A$-based \metal. There is a clear trend that gives larger \agessp\ discrepancies at larger \metalssp.}
\label{newfind}
\end{figure}

Fig.\ref{differences} and Fig.\ref{totaldeltas} show another interesting feature: the largest difference between \age$\rm_{H\beta}$ and \age$\rm_{H\gamma_A}$ in Fig.\ref{differences} is small compared to the one in Fig.\ref{totaldeltas}. Recall that Fig.\ref{differences} is relative to CSP's where both SSP$_1$ and SSP$_2$ have solar chemical composition, while Fig.\ref{totaldeltas} represents the whole BC03 sample, with \metal\ growing significantly above solar. The two figures suggest that \metal\ plays an important role with respect to the difference \age$\rm_{H\beta}$--\age$\rm_{H\gamma_A}$. Fig.\ref{newfind} shows that indeed high (and therefore more easily detectable) differences in \agessp\ occurr only at high metallicities. Similar plots hold for the difference in \metal\ and \abund. We therefore find that the use of more than one Balmer-line index can reveal the presence of a young stellar populations, but that this is possible only for a small range of \age$_2$ and $\mu$ and depends also on the metallicity of the populations.

We would like to remind the reader that a disagreement between Balmer-line-based SSP-equivalent parameters, in principle revealing the presence of a young stellar component, could also result from the approach used when analysing the data. In particular, it is important to remember that different Balmer-line indices respond differently to variations in \abund. Thomas et al.\ (2004) and Thomas \& Davies (2006) pointed out that this causes a discrepancy between SSP-equivalent parameters determined from different Balmer-line indices when using as a comparison models with solar \abund\ only. This effect could mimic the presence of a young stellar component. However, no such problem should occur when using models that account properly for \abund\  variations, as was done here.

Another delicate point when using Balmer-line indices is their increase caused by hot star populations like blue horizontal branch stars or blue stragglers (Maraston \& Thomas 2000; Trager et al.\ 2005). Although this is not an issue for the present study, where we do not explore the metal-poor regime at which these stars are expected to be found, this problem should always be kept into consideration when dealing with real data.

\section{Conclusions}

We have built two datasets of composite stellar populations (CSP) using Bruzual \& Charlot (2003) and Worthey (1994) models. Each CSP model in the datasets consists of an old single-burst stellar population (SSP$_1$) and a younger, less massive one (SSP$_2$). We have investigated how the SSP-equivalent parameters determined by measuring the Lick/IDS line-strength indices of the CSP's depend on the stellar parameters of SSP$_1$ and SSP$_2$. By means of covariance coefficients we have found that, regardless of the particular stellar populations models used in building the CSP's and of the Balmer-line index used for the analysis: \agessp, the SSP-equivalent age, depends primarily on \age$_2$, the age of the young population, and $\mu$, the mass fraction between the two populations; variations in \metalssp, the SSP-equivalent metallicity, are mostly driven by variations in \metal$_1$, the metallicity of the old population; and \abundssp\, the SSP-equivalent abundance ratio, depends mainly on \abund$_1$, the abundance ratio of the old population.

Furthermore, we have found that \metalssp\ and \abundssp\ track quite closely the $V$-band luminosity-weighted  metallicity and abundance ratio (\metall\ and \abundl) except in case of very young (and significantly massive) SSP$_2$. On the other hand, \agessp\ does not follow \agel, being strongly biased towards the the age \age$_2$ of the young population. The SSP-equivalent age \agessp\ is simply a Balmer-line-weighted age \emph{and should not be interpreted as the time passed since the formation of most of the stars in a galaxy.}

Finally, as found by Schiavon et al.\ (2004), using more than one Balmer-line index can reveal the presence of a young stellar component on top of an old one. In this case, SSP-equivalent parameters derived from different Balmer-line index give discrepant results. This is true however only for values of $\mu$ between 1\% and 10\% and for \age$_2\leq$2.5 Gyr assuming typical errors on \agessp, \metalssp\ and \abundssp. Furthermore, these discrepancies are higher at supersolar \metalssp. Finally, this method does not appear to break the degeneracy between the age and the mass fraction of the young population, especially when considering the size of the typical error bars. In this respect, what is really needed is an age-sensitive index dependent on the age of the old stellar population  (i.e., RGB stars), to be used in combination with Balmer-line indices.

\section*{Acknowledgements}

The authors would like to thank the referee, Daniel Thomas, for useful comments that helped clarify important points of the discussion. We also thank Guy Worthey for providing new index responses to abundance ratio variations in advance of publication.

\label{lastpage}


\begin{thebibliography}{}

\bibitem[]{} Bruzual A.G., Charlot S., 2003, MNRAS, 344, 1000

\bibitem[]{} Burstein D., Faber S.M., Gaskell C.M., Krumm, N., 1984, ApJ, 287, 586	

\bibitem[]{} Caldwell, N., Rose, J.A., Concannon, K.D., 2003, AJ, 125, 2891

\bibitem[]{} Clemens, M.S., Bressan, A., Nikolic, B., Alexander, P., Annibali, F., Rampazzo, R., 2006, MNRAS, 370, 702	

\bibitem[]{} Denicol\`{o}, G., Terlevich, R., Terlevich, E., Forbes, D.A., Terlevich, A., 2005, MNRAS, 358, 813	

\bibitem[]{} Faber, S.M., 1973, ApJ, 179, 731

\bibitem[]{} Jeong, H., Bureau, M., Yi, S.K., Krajnovi\'{c}, D., Davies, R.L., 2006, MNRAS, submitted, astro-ph/0608212

\bibitem[]{} Leonardi, A.J., Worthey, G., 2000, ApJ, 534, 650

\bibitem[]{} Maraston C., Thomas D., 2000, ApJ, 541, 126

\bibitem[]{} O'Connell, R.W., 1986, PASP, 98, 163

\bibitem[]{} Schiavon, R., Caldwell, N., Rose, J.A., 2004, AJ, 127, 1513

\bibitem[]{} Thomas D., Davies R.L., 2006, MNRAS, 366, 510

\bibitem[]{} Thomas D., Maraston C., Bender R., 2003, MNRAS, 339, 897

\bibitem[]{} Thomas D., Maraston C., Korn A., MNRAS, 351L, 19

\bibitem[]{} Thomas, D., Maraston, C., Bender, R., Mendes de Oliveira, C., 2005, ApJ, 621, 673	

\bibitem[]{} Trager, S.C., Faber, S.M., Worthey, G., Gonz{\'a}lez, J.J.\ 2000a, AJ, 119, 1645

\bibitem[]{} Trager, S.C., Faber, S.M., Worthey, G., Gonz{\'a}lez, J.J.\ 2000b, AJ, 120, 165

\bibitem[]{} Trager S.C., Worthey G., Faber S.M., Dressler A., 2005, MNRAS, 362, 2

\bibitem[]{} Trager, S.C., Faber, S.M., Dressler, A., 2006, MNRAS, submitted

\bibitem[]{} Vazdekis, A., 1999, ApJ, 513, 224

\bibitem[]{} Worthey, G., 1994, ApJS, 95, 107

\bibitem[]{} Worthey, G., Ottaviani, D.L., 1997, ApJS, 111, 377

\bibitem[]{} Worthey, G., Faber, S.M., Gonz{\'a}lez, J.J., Burstein, D., 1994, ApJS, 94, 687

\bibitem[]{} Yi, S.K. et al.\  2005, ApJ, 619L, 111

\end{thebibliography}
\end{document}